\documentclass{llncs}
\usepackage[dvips]{graphicx}
\usepackage[normal]{subfigure}
\usepackage{verbatim} 

\begin{document}

\title{DWEB: A Data Warehouse\\ Engineering Benchmark}

\author{J\'{e}r\^{o}me Darmont \and Fadila Bentayeb \and Omar Boussa\"{i}d}

\institute{ERIC, University of Lyon 2\\
5 av. Pierre Mend\`{e}s-France\\
69676 Bron Cedex\\
France\\
\{jdarmont$|$boussaid$|$bentayeb\}@eric.univ-lyon2.fr}


\maketitle

\begin{abstract}
Data warehouse architectural choices and optimization techniques are critical to decision support query performance. To facilitate these choices, the performance of the designed data warehouse must be assessed. This is usually done with the help of benchmarks, which can either help system users comparing the performances of different systems, or help system engineers testing the effect of various design choices. While the TPC standard decision support benchmarks address the first point, they are not tuneable enough to address the second one and fail to model different data warehouse schemas. By contrast, our Data Warehouse Engineering Benchmark (DWEB) allows to generate various ad-hoc synthetic data warehouses and workloads. DWEB is fully parameterized to fulfill data warehouse design needs. However, two levels of parameterization keep it relatively easy to tune. Finally, DWEB is implemented as a Java free software that can be interfaced with most existing relational database management systems. A sample usage of DWEB is also provided in this paper.
\end{abstract}

\section{Introduction}
\label{sec:introduction}

When designing a data warehouse, choosing an architecture is crucial. Since it is very dependant on the domain of application and the analysis objectives that are selected for decision support, different solutions are possible. In the ROLAP (Relational OLAP) environment we consider, the most popular solutions are by far star, snowflake, and constellation schemas~\cite{INM02,KIM02}, and other modeling possibilities might exist. This choice of architecture is not neutral: it always has advantages and drawbacks and greatly influences the response time of decision support queries. 
Once the architecture is selected, various optimization techniques such as indexing or materialized views further influence querying and refreshing performance. Again, it is a matter of trade-off between the improvement brought by a given technique and its overhead in terms of maintenance time and additional disk space; and also between different optimization techniques that may cohabit.
To help users make these critical choices of architecture and optimization techniques, the performance of the designed data warehouse needs to be assessed. However, evaluating data warehousing and decision support technologies is an intricate task. Though pertinent, general advice is available, notably on-line~\cite{GRE04a,PEN03}, more quantitative elements regarding sheer performance are scarce. Thus, we propose in this paper a data warehouse benchmark we named DWEB (the \emph{Data Warehouse Engineering Benchmark}).
Different goals may be achieved by using a benchmark:
	(1) compare the performances of various systems in a given set of experimental conditions (users);
	(2) evaluate the impact of architectural choices or optimisation techniques on the performances of one given system
	(system designers).
The Transaction Processing Performance Council (TPC), 
a non-profit organization, defines standard benchmarks and publishes objective and verifiable performance evaluations to the industry. Out of the TPC, few decision support benchmarks have been designed. Some do exist, but their specification is not fully published~\cite{DEM95}. Some others are not available any more, such as the OLAP APB-1 benchmark that was issued in the late nineties by the OLAP council, an organization whose web site does not exist any more.

The TPC benchmarks mainly aim at the first benchmarking goal we identified. However, the database schema of TPC benchmarks TPC-H~\cite{TPC03a} and 
 TPC-R~\cite{TPC03b} 
is a classical \emph{product-order-supplier} model, and not a typical data warehouse schema such as a star schema and its derivatives. Furthermore, their workload, though decision-oriented, does not include explicit OLAP (On-Line Analytical Processing) queries either, and they do not address specific warehousing issues such as the ETL (Extract, Transform, Load) process. These benchmarks are indeed implicitely considered obsolete by the TPC that has issued some specifications for their successor: TPC-DS~\cite{POE02}. However, TPC-DS has been under development for three years now and is not completed yet. 
Furthermore, although the TPC decision support benchmarks are scaleable according to Gray's definition~\cite{GRA93},
their schema is fixed. 
It must be used ``as is''. Different ad-hoc configurations are not possible. 
There is only one parameter to define the database, the Scale Factor ($SF$), which sets up its size (from 1 to 100,000~GB). The user cannot control the size of the dimensions and the fact tables separately, for instance. Finally, 
the user has no control on the workload's definition. The TPC benchmarks are thus not well adapted to evaluate the impact of architectural choices or optimisation techniques on global performance. 
For these reasons, we decided to design a full data warehouse synthetic benchmark that would be able to model various ad-hoc configurations of database (modeled as star, snowflake, or constellation schemas) and workload, while being simpler to develop than TPC-DS. We mainly seek to fulfill engineering needs (second benchmarking objective). 

This paper presents an overview the DWEB benchmark. First, we present our benchmark's database (metaschema, parameterization, and instiantiation into an actual data warehouse) in Section~\ref{sec:dweb-database}. Then, we present the benchmark's workload (query model, parameterization, and workload generation) in Section~\ref{sec:dweb-workload}. We illustrate how our benchmark can be used in Section~\ref{sec:sample-usage} and finally conclude this paper and provide future research directions in Section~\ref{sec:conclusion}.

\section{DWEB database}
\label{sec:dweb-database}

\subsection{Schema}

\begin{figure*}[hbt]
	\centering
	\includegraphics[width=10cm]{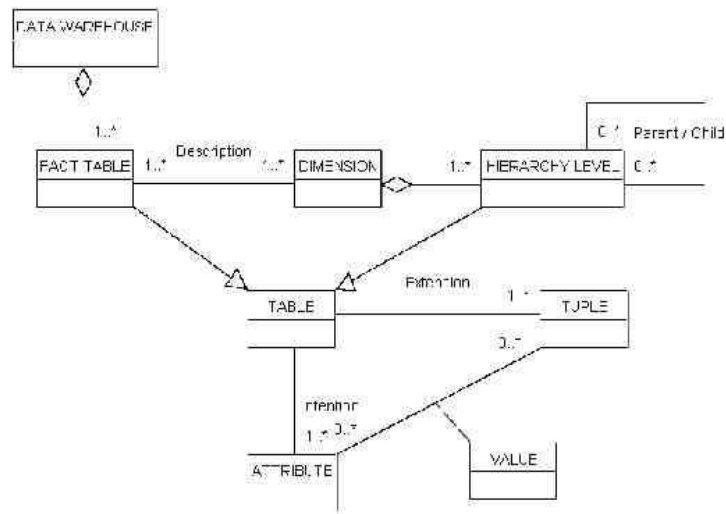}
	\caption{DWEB data warehouse metaschema}
	\label{fig:metamodel}
\end{figure*}

Our design objective for DWEB is to be able to model the different kinds of data warehouse architectures that are popular within a
ROLAP environment:
	classical star schemas;
	snowflake schemas with hierarchical dimensions; and
	constellation schemas with multiple fact tables and shared dimensions.
To achieve this goal, we propose a data warehouse metamodel (represented as a UML class diagram in
Figure~\ref{fig:metamodel}) that can be instantiated
into these different schemas. We view this metamodel as a middle ground between the multidimensional 
metamodel from the Common Warehouse Metamodel (CWM 
\cite{OMG03}) and the eventual benchmark model. Our metamodel
is actually an instance of the CWM metamodel, which could be qualified as a meta-metamodel in our context.

Our metamodel is relatively simple, but it is sufficient to model the data warehouse schemas we aim at (star, snowflake, and constellation schemas). Its upper part describes a data warehouse (or a datamart, if a datamart is viewed as a small, dedicated data warehouse) as constituted of one or several fact tables that are each described by several dimensions. Each dimension may also describe several fact tables (shared dimensions). Each dimension may be constituted of one or several hierarchies made of different levels. There can be only one level if the dimension is not a hierarchy.
Both fact tables and dimension hierarchy levels are relational tables, which are modeled in the lower part of Figure~\ref{fig:metamodel}. Classically, a table or relation is defined in intention by its attributes and in extension by its tuples or rows. At the intersection of a given attribute and a given tuple lies the value of this attribute in this tuple.

\subsection{Parameterization}
\label{db-parameters}

DWEB's database parameters help users selecting the data warehouse architecture they need in a given context.
The main difficulty in producing a data warehouse schema is parameterizing the instantiation of the metaschema. We indeed try to meet the four key criteria that make a ``good'' benchmark, as defined by Gray~\cite{GRA93}:
	\emph{relevance:} the benchmark must answer our engineering needs 
	\emph{portability:} the benchmark must be easy to implement on different systems;
	\emph{scalability:} it must be possible to benchmark small and large databases, and to scale up the benchmark;
	and \emph{simplicity:} the benchmark must be understandable, otherwise it will not be credible nor used.
Relevance and simplicity are clearly two orthogonal goals. Introducing too few parameters reduces the model's expressiveness, while introducing too many parameters makes it difficult to apprehend by potential users. Furthermore, few of these parameters are likely to be used in practice. In parallel, the generation complexity of the instantiated schema must be mastered.
To solve this dilemna, 
we propose to divide the parameter set into two subsets.
	The first subset of so-called low-level parameters allows an advanced user to control everything about the data warehouse generation. 
	However, the number of low-level parameters can increase dramatically when the schema gets larger. For instance, if there are several fact tables, all their characteristics, including dimensions and their own characteristics, must be defined for each fact table.
	Thus, we designed a layer above with much fewer parameters that may be easily understood and set up (Table~\ref{tab:HighLevelParameters}). More precisely, these high-level parameters are average values for the low-level parameters. At database generation time, the high-level parameters are exploited by random functions (following a gaussian distribution) to automatically set up the low-level parameters. Finally, unlike the number of low-level parameters, the number of high-level parameters always remains constant and reasonable (less than ten parameters).
Users may choose to set up either the full set of low-level parameters, or only the high-level parameters, for which we propose default values that correspond to a snowflake schema. These parameters control both schema and data generation.

\begin{table*}[hbt]
	\centering
		\begin{tabular}{|l|l|c|}
			\hline 
			\small{\textbf{Parameter name}} & \small{\textbf{Meaning}} & \small{\textbf{Def. val.}} \\
			\hline \hline
			\small{$AVG\_NB\_FT$} & \small{Average number of fact tables} & \small{1} \\
			\hline
			\small{$AVG\_NB\_DIM$} & \small{Average number of dimensions per fact table} & \small{5} \\
			\hline
			\small{$AVG\_TOT\_NB\_DIM$} & \small{Average total number of dimensions} & \small{5} \\
			\hline
			\small{$AVG\_NB\_MEAS$} & \small{Average number of measures in fact tables} & \small{5} \\
			\hline
			\small{$AVG\_DENSITY$} & \small{Average density rate in fact tables} & \small{0.6} \\
			\hline
			\small{$AVG\_NB\_LEVELS$} & \small{Average number of hierarchy levels in dimensions} & \small{3} \\
			\hline
			\small{$AVG\_NB\_ATT$} & \small{Average number of attributes in hierarchy levels} & \small{5} \\
			\hline
			\small{$AVG\_HHLEVEL\_SIZE$} & \small{Average number of tuples in highest hierarchy levels} & \small{10} \\
			\hline
			\small{$DIM\_SFACTOR$} & \small{Average size scale factor within hierarchy levels} & \small{10} \\
			\hline			
		\end{tabular}
	\caption{DWEB warehouse high-level parameters}
	\label{tab:HighLevelParameters}
\end{table*}

Note that 
the cardinal of a fact table is usually lower or equal to the product of its dimensions' cardinals. This is why we introduce the notion of density. A density rate of one indicates that all the possible combinations of the dimension primary keys are present in the fact table. When the density rate decreases, we progressively eliminate some of these combinations.
This parameter helps controlling the size of the fact table, independantly of the size of its dimensions.
Furthermore, within a dimension, a given hierarchy level normally has a greater cardinality than the next level. For example, in a \emph{town-region-country} hierarchy, the number of towns must be greater than the number of regions, which must be in turn greater than the number of countries. There is also often a significant scale factor between these cardinalities (e.g., one thousand towns, one hundred regions, ten countries). Hence, we model the cardinality of hierarchy levels by assigning a ``starting'' cardinality to the highest level in the hierarchy ($HHLEVEL\_SIZE$), and then by multiplying it by a predefined scale factor ($DIM\_SFACTOR$) for each lower-level hierarchy.
Finally, since some of DWEB's parameters might sound abstract, the data warehouse global size (in megabytes) is assessed at generation time so that users retain full control over it and may adjust the parameters to better represent the kind of warehouse they need.

\subsection{Generation algorithm}
\label{sec:genalgorithm}

The instantiation of the DWEB metaschema into an actual benchmark schema is done in two steps: (1) build the dimensions; (2) build the fact tables.
Due to space constraints, the pseudo-code for these two steps is not provided here, but it is available on-line \cite{dweb05full}.
Each of these steps is further subdivided, for each dimension and each fact table, into generating its intention and extension. In addition, hierarchies of dimensions are managed. 

\section{DWEB workload}
\label{sec:dweb-workload}

In a data warehouse benchmark, the workload may be subdivided into
	a load of decision support queries (mostly OLAP queries)
	and the ETL (data generation and maintenance) process.
To design DWEB's workload, we inspire both from TPC-DS' workload definition 
and information regarding data warehouse performance from other sources~\cite{BMC00,GRE04b}. However, TPC-DS' workload is very elaborate and sometimes confusing. Its reporting, ad-hoc decision support and OLAP query classes are very similar, for instance, but none of them include any specific OLAP operator such as Cube or Rollup. Since we want to meet Gray's simplicity criterion, we propose a simpler workload. Furthermore, we also have to design a workload that is consistent with the variable nature of the DWEB data warehouses.
We also, in a first step, mainly focus on the definition of a query model. Modeling the full ETL process is a complex task that we postpone for now. We consider that the current DWEB specifications provide a raw loading evaluation framework. The DWEB database may indeed be generated into flat files, and then loaded into a data warehouse using the ETL tools provided by the system.

\subsection{Query model}
\label{sec:query-model}

The DWEB workload models two different classes of queries:
  purely decision-oriented queries involving common OLAP operations, such as cube, roll-up, drill down and slice and dice;
  and extraction queries (simple join queries).
We define our generic query model as a grammar that is a subset of the SQL-99 standard. 
Due to space constraints, this query model is only available on-line \cite{dweb05full}.

\subsection{Parameterization}
\label{sec:query-param}

DWEB's workload parameters help users tailoring the benchmark's load, which is also dependent from the warehouse schema, to their needs.
Just like DWEB's database paramameter set, 
DWEB's workload parameter set (Table~\ref{tab:workload-parameters}) has been designed with Gray's simplicity criterion in mind. These parameters determine how the query model 
is instantiated.
These parameters help defining the workload's size and complexity, by setting up the proportion of complex OLAP queries (i.e., the class of queries) in the workload, the number of aggregation operations, the presence of a Having clause in the query, or the number of subsequent drill down operations.
Here, we have only a limited number of high-level parameters 
Indeed, it cannot be envisaged to dive further into detail if the workload is as large as several hundred queries, which is quite typical. 
Note that $NB\_Q$ is only an \textit{approximate} number of queries because the number of drill down operations after an OLAP query may vary. Hence we can stop generating queries only when we actually have generated as many or more queries than $NB\_Q$.

\begin{table*}[hbt]
	\centering
		\begin{tabular}{|l|l|c|}
			\hline 
			\small{\textbf{Parameter name}} & \small{\textbf{Meaning}} & \small{\textbf{Def. val.}} \\
			\hline \hline
			\small{$NB\_Q$} & \small{Approximate number of queries in the workload} & \small{100} \\
			\hline
			\small{$AVG\_NB\_ATT$} & \small{Average number of selected attributes in a query} & \small{5} \\
			\hline
			\small{$AVG\_NB\_RESTR$} & \small{Average number of restrictions in the query} & \small{3} \\
			\hline
			\small{$PROB\_OLAP$} & \small{Probability that the query type is OLAP} & \small{0.9} \\
			\hline
			\small{$PROB\_EXTRACT$} & \small{Probability that the query is an extraction query} & \tiny{$1-P\_OLAP$} \\						
			\hline
			\small{$AVG\_NB\_AGGREG$} & \small{Average number of aggregations in an OLAP query} & \small{3} \\
			\hline
			\small{$PROB\_CUBE$} & \small{Probability of an OLAP query to use the Cube operator} & \small{0.3} \\
			\hline
			\small{$PROB\_ROLLUP$} & \small{Probability of an OLAP query to use the Rollup operator} & \tiny{$1-P\_CUBE$} \\
			\hline
			\small{$PROB\_HAVING$} & \small{Probability of an OLAP query to include an Having clause} & \small{0.2} \\
			\hline			
			\small{$AVG\_NB\_DD$} & \small{Average number of drill downs after an OLAP query} & \small{3} \\
			\hline
		\end{tabular}
	\caption{DWEB workload parameters}
	\label{tab:workload-parameters}
\end{table*}

\subsection{Generation algorithm}

Due to space constraints, the pseudo-code of DWEB's workload generation algorithm is only available on-line \cite{dweb05full}.
However, its principle follows.
The algorithm's purpose is to generate a set of SQL-99 queries that can be directly executed on the synthetic data warehouse defined in Section~\ref{sec:dweb-database}. It is subdivided into two steps:
	(1) generate an initial query that may either be an OLAP or an extraction (join) query;
  (2) if the initial query is an OLAP query, execute a certain number of drill down operations based on the first OLAP query. More precisely, each time a drill down is performed, an attribute from a lower level of dimension hierarchy is added to the attribute clause of the previous query.
Step~1 is further subdivided into three substeps:
	(1) the Select, From, and Where clauses of a query are generated simultaneously by randomly selecting a fact table and dimensions, including a hierarchy level within a given dimension hierarchy;
	(2) the Where clause is supplemented with additional conditions;
	(3) eventually, it is decided whether the query is an OLAP query or an extraction query. In the second case, the query is complete. In the first case, aggregate functions applied to measures of the fact table are added in the query, as well as a Group by clause that may include either the Cube or the Rollup operator. A Having clause may optionally be added in too. The aggregate function we apply on measures is always Sum since it is the most common aggregate in cubes. Furthermore, other aggregate functions bear similar time complexities, so they would not bring in any more insight in a performance study.

\section{Sample usage of DWEB}
\label{sec:sample-usage}

In order to illustrate one possible usage for DWEB, we tested the efficiency of bitmap join indices, which are well suited to the data warehouse environment, 
on decision support queries under Oracle. 
The aim of this particular example is to compare the execution time of a given workload on a given data warehouse, with and without using bitmap join indices.

First, we generated a data warehouse modeled as a snowflake schema. 
This schema is organized around one fact table that is described by five dimensions, each bearing two to three hierarchy levels.
The fact table contains about 140,000 tuples, the dimension hierarchy levels about ten tuples on an average, for a global size of about 4~MB 
(this is a voluntarily small example and not a full-scale test).
We applied different workloads on this data warehouse. $Workload \#1$ is a typical DWEB workload 
constituted of fifty queries.
10\% of these queries are extraction (join) queries and the rest are decision support queries involving OLAP operators (Cube and Rollup). In $Workload \#1$, we limited the queries to the dimensions' lowest hierarchy levels, i.e., to the star schema constituted of the fact table and the ``closest'' hierarchy levels.
$Workload \#2$ is similar to $Workload \#1$, but it is extended with drill down operations that scan the dimensions' full hierarchies (from the highest level to the lowest level). Thus, this workload exploits the whole snowflake schema.
To evaluate the efficiency of bitmap join indices, we timed the execution of these two workloads on our test data warehouse (response time is our only performance metric for now), first with no index, and then by forcing the use of five bitmap join indices defined on the five dimensions (for the lowest hierarchy levels in $Workload \#1$ and for the whole hierarchies in $Workload \#2$). To flaten any response time variation in these experiments, we replicated each test ten times and computed the average response times. We made sure \emph{a posteriori} that the standard deviation was close to zero. These tests have been executed on a PC with a Celeron~900 processor, 128~MB of RAM, an IDE hard drive, and running Windows XP Professional and Oracle 9i.

\begin{figure}[hbt]
	\centering
	\subfigure{\includegraphics[width=6.cm]{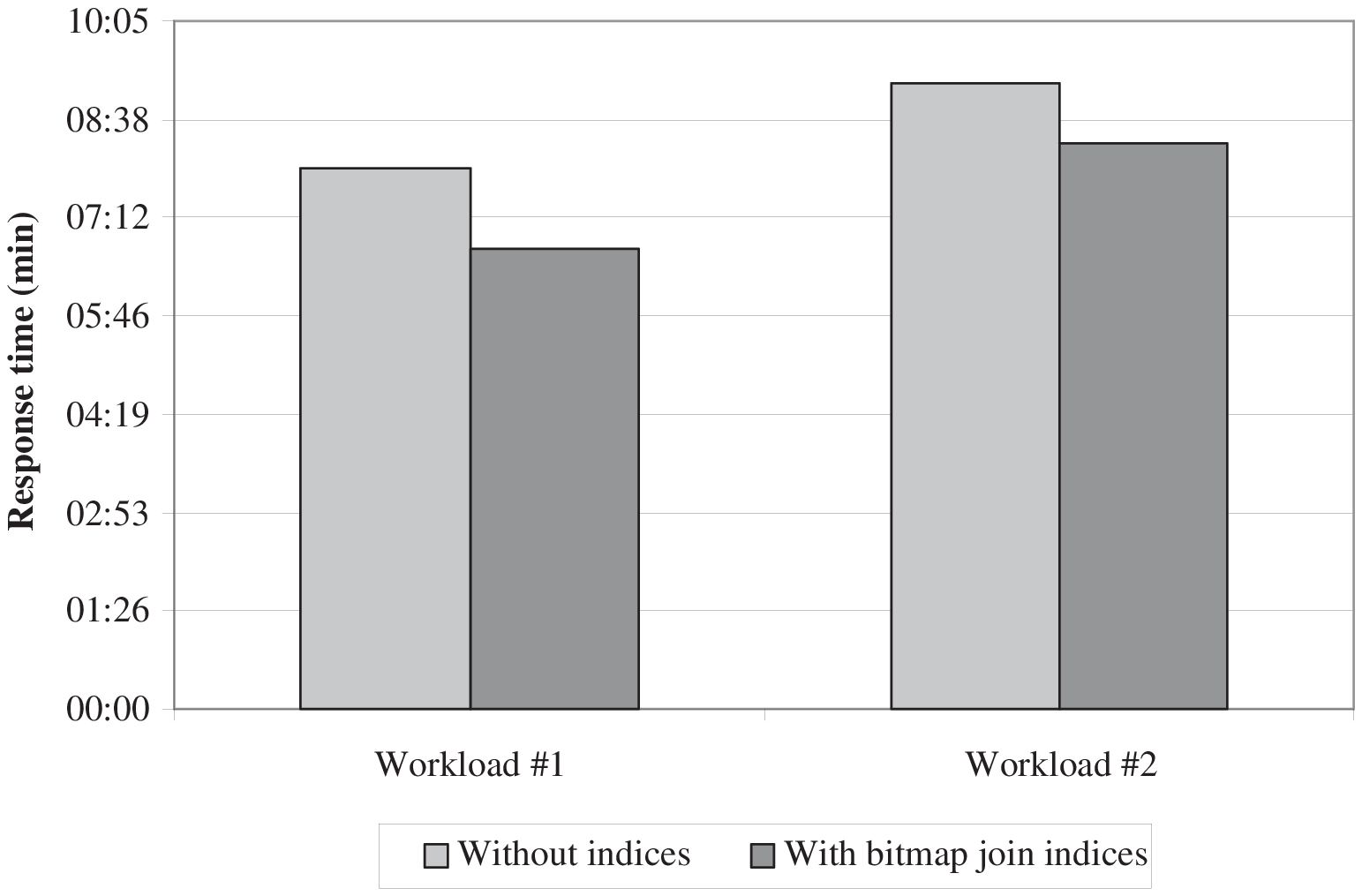}}
	\subfigure{\includegraphics[width=6.cm]{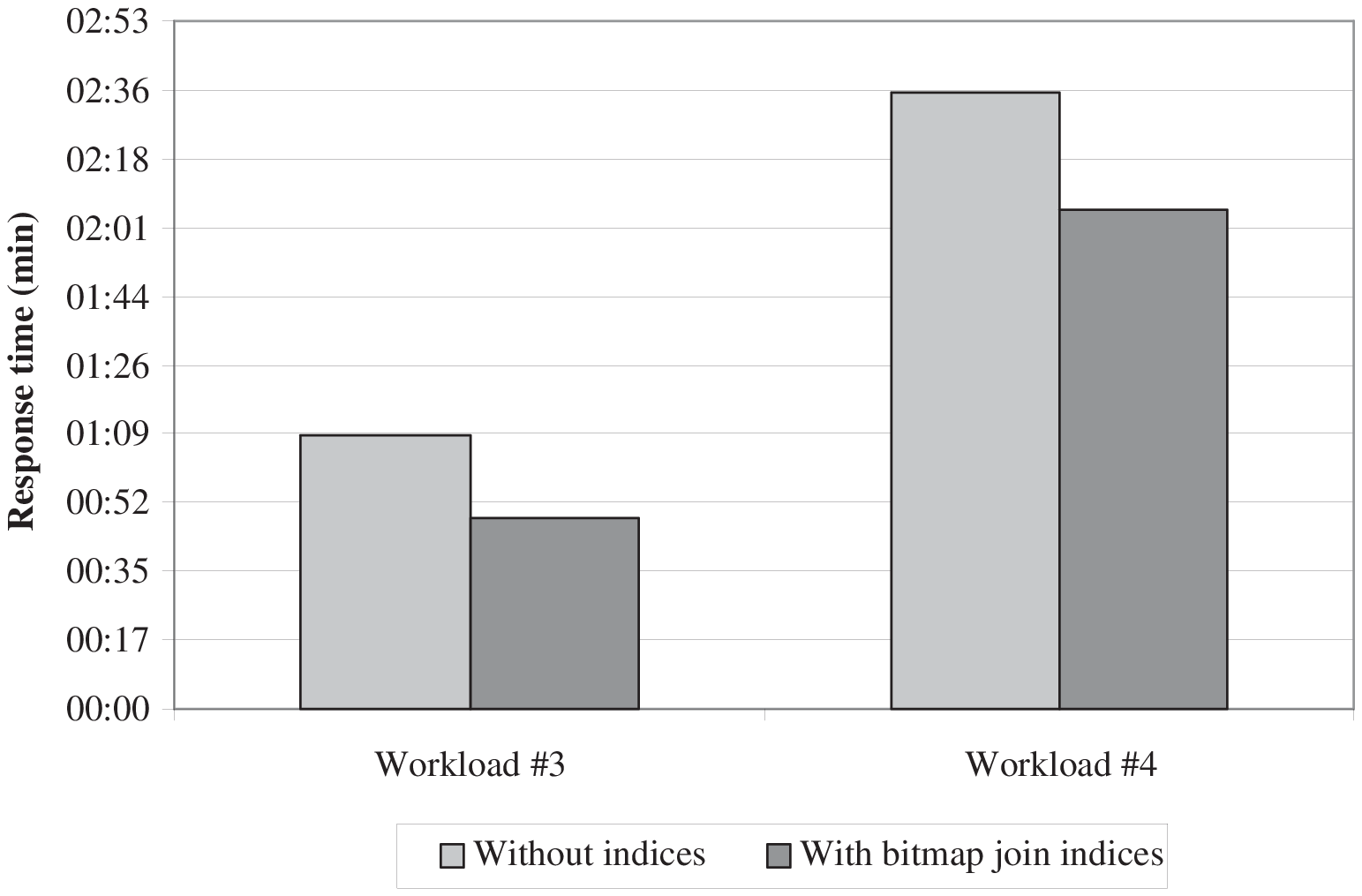}}
	\caption{Test results}
	\label{fig:tests}
\end{figure}

The left-hand graph on Figure~\ref{fig:tests} represents the average response time achieved for $Workload \#1$ and $\#2$, with and without bitmap join indices, respectively. It shows a gain in performance of 15\% for $Workload \#1$, and 9.4\% for $Workload \#2$. This decrease in efficiency was expected, since the drill down operations added in $Workload \#2$ are costly and need to access the data (bitmap join indices alone cannot answer such queries). However, the overall performance improvement we achieved was not as good as we expected. 
We formulated the hypothesis that the extraction queries, which are costly joins and need to access the data too, were not fully benefiting from the bitmap join indices. To confirm this hypothesis, we generated two new workloads, $Workload$ $\#3$ and $\#4$. They are actually almost identical to $Workload \#1$ and $Workload \#2$, respectively, but do not include any extraction (join) queries. Then, we repeated our experiment following the same protocol.
The right-hand graph on Figure~\ref{fig:tests} represents the average response time achieved for $Workload \#3$ and $\#4$, with and without bitmap join indices, respectively. This time, we obtained similar results than in our previous experiment (in trend): response time clearly increases when drill down operations are included into the workload. However, response time is now much better and the gain in performance is 30.9\% for $Workload \#3$, and 19.2\% for $Workload \#4$.

In conclusion, we want to point out that
these experiments are not very significant \emph{per se}, and do not do justice to Oracle. 
However, we illustrated how DWEB could be used for performance evaluation purposes. These experiments could also be seen as a (very basic) performance comparison between two different data warehouse architectures (star schema and snowflake schema). Our results indeed conform to the well-known fact that introducing hierarchies into a star schema induces more join operations in the decision support queries, and hence degrade their response time. Finally, we were also able to witness the impact of costly join operations on a data warehouse structure that is not properly indexed to answer such queries. 

\section{Conclusion and perspectives}
\label{sec:conclusion}

We proposed in this paper a new data warehouse benchmark called DWEB (the \emph{Data Warehouse Engineering Benchmark}) that is aimed at
helping data warehouse designers to choose between alternate warehouse architectures and performance optimization techniques. 
When designing DWEB, we tried to grant it the characteristics that make up a ``good'' benchmark according to Gray: relevance, portability, scalability, and simplicity. To make DWEB relevant for evaluating the performance of data warehouses in an engineering context, we designed it to generate different data warehouse schemas (namely star, snowflake and constellation schemas) and workloads. 
Note that the database schema of TPC-DS, the future standard data warehouse benchmark currently developped by the TPC, can be modeled with DWEB. In addition, though DWEB's workload is not currently as elaborate as TPC-DS's, it is also much easier to implement. It will be important to fully include the ETL process into our workload, though, and the specifications of TPCD-DS and some other existing studies~\cite{LAB98} might help us.
We now need to further test DWEB's relevance on real cases. To achieve this goal, we plan to compare the efficiency of various index and materialized view selection techniques.
We also made DWEB very tuneable to reach both the relevance and scalability objectives. However, too many parameters make the benchmark complex to use and contradict the simplicity requirement. Though it is impossible to achieve both a high simplicity and a high relevance and scalability, we introduced a layer of high-level parameters that are both simpler than the potentially numerous low-level parameters, and in reduced and constant number. DWEB might not be qualified as a simple benchmark, but our objective was to keep its complexity as low as possible.
Finally, portability was achieved through a Java implementation. 
DWEB's latest version is freely available on-line~\cite{DWEB05}. Finally, we also illustrated with a practical case how DWEB can be used.

This work opens up many perspectives for developing and enhancing DWEB. In this paper, we assumed an execution protocol and performance metrics were easy to define for DWEB (e.g., using TPC-DS' as a base) and focused on the benchmark's database and workload model. 
A more elaborate execution protocol must be designed, especially since two executions of DWEB using the same parameters produce different data warehouses and workloads. This is interesting when, for instance, one optimization technique needs to be tested against many databases. However, note that it is also possible to save a given warehouse and its associated workload to run tests on different systems and/or with various optimization techniques.
Defining sound metrics (beside response time) would also improve DWEB's usefulness. In this area, we could inspire from metrics designed to measure the quality of data warehouse conceptual models~\cite{serrano04a}. 
We are also currently working on warehousing complex, non-standard data (including multimedia data, for instance). Such data may be stored as XML documents. Thus, we also plan a ``complex data'' extension of DWEB that would take into account the advances in XML warehousing.
Finally, more experiments with DWEB should also help us 
propose sounder default parameter values. We also encourage other people
to report on their own experiments.

\bibliographystyle{abbrv}
\bibliography{ds_benchmarks}

\end{document}